\documentclass[aps,prb,twocolumn,preprintnumbers,amsmath,amssymb,]{revtex4}
\usepackage{graphicx}
\begin{document}
\title{Physical properties of  (Mn$_{0.85}$Fe$_{0.15}$)Si along the critical trajectory}

\author{A.E. Petrova}
\author{S.M. Stishov}
\email{sergei@hppi.troitsk.ru}
\affiliation{Institute for High Pressure Physics of RAS, Troitsk, Moscow, Russia}
\author{Dirk Menzel}
\affiliation{Institut f\"{u}r Physik der Kondensierten Materie, Technische Universit\"{a}t Braunschweig, D-38106 Braunschweig, Germany}

\begin{abstract}
We report results of studying the magnetization, specific heat and thermal expansion  of a single crystal with nominal composition (Mn$_{0.85}$Fe$_{0.15}$)Si. We found no thermodynamic evidences in favor of a second order phase transition in the 15\% Fe substituted MnSi. The trajectory corresponding to the present composition of (MnFe)Si is a critical one, i.e. approaching quantum critical point at lowering temperature, but some properties may feel the cloud of helical fluctuations bordering the phase transition line. 
\end{abstract}
\maketitle

\section{Introduction}
An evolution of the magnetic phase transition in the helical magnet MnSi at high pressure is reported in a number of publications~\cite{1,2,3}. It became clear that the phase transition temperature decreased with pressure and practically reached the zero value at $\sim$15 kbar. However a nature of this transition at zero temperature and high pressure is still a subject of controversial interpretations. Early it was claimed an existence of tricritical point on the phase transition line that might result in a first order phase transition in MnSi at low temperatures~\cite{4}. The latter would prevent an existence of quantum critical point in MnSi.  This view was seemingly supported by the volume measurements at the phase transition in MnSi~\cite{5,6}. However, this idea was disputed in papers~\cite{7,8}, where stated that the observed volume anomaly at the phase transitions in MnSi at low temperatures was simply the slightly narrowing anomaly clearly seen at elevated temperatures. On the other hand some experimental works and the recent Monte-Carlo calculations may indicate a strong influence of inhomogeneous stress arising at high pressures and low temperatures on characteristics of phase transitions that could make any experimental data not entirely conclusive~\cite{7,8,9}.

In this situation it would be appealing to use a different approach to discover a quantum criticality in MnSi, for instance, making use doping as a controlling parameter. Indeed, it became known that doping MnSi with Fe and Co decreases a temperature of the magnetic phase transitions and finally completely suppress the transitions at some critical concentrations of the dopants. In case of Fe doping a critical concentration consist about 15\% (actually different estimates vary from 0.10 to 0.19)~\cite{10,11,13}. 

Actually, the general belief that the concentration of the dopant added to the batch will be the same in the grown crystal is incorrect. One needs to preform  chemical and x-ray analysises to make a certain conclusion about the real composition of material. Anyway there are some evidences (non Fermi liquid resistivity, logarithmic divergence of specific heat) that indeed the quantum critical point occurs in (MnFe)Si in the vicinity of iron concentration 0.15\% at ambient pressure. However, in the recent publication it is claimed that (Mn$_{0.85}$Fe$_{0.15}$)Si experiences a second order phase transition at the pressure range to $\sim$0-23 kbar, therefore placing the quantum critical point in this material at high pressure~\cite{14}.

To this end it seems appropriate to take another look at the situation. 
We report here results  of study of a single crystal with nominal composition Mn$_{0.85}$Fe$_{0.15}$)Si.  The sample was prepared from the ingot obtained by premelting of Mn (purity 99.99\%, Chempur), Fe (purity 99.98\%, Alfa Aesar), and Si ($\rho_n$=300 Ohm cm, $\rho_p$=3000 Ohm cm) under argon atmosphere in a single arc oven, then a single crystal was grown using the triarc Czochralski technique. 
The electron-probe microanalysis shows that real composition is (Mn$_{0.795}$Fe$_{0.147}$)$_{47.1}$Si$_{52.9}$, which indicates some deviations from the stoichiometric chemical compositions common to the silicide compounds. But hereafter we will call the sample under study as (MnFe)Si.

The lattice parameter of the sample appeared to be a=4.5462\AA. Note that the lattice parameter of pure MnSi is somewhat higher and equal to a=4.5598 \AA. This implies that iron plays a role of some sort of pressure agent. Let’s estimate what pressure is needed to compress pure MnSi to the volume corresponding to the lattice parameter of the material under study. We use a simple linear expression of the form $P=K\dfrac{\Delta V}{V}$, where $P$-pressure, $K=-V(\frac{dP}{dV})_T$- bulk modulus, $\dfrac{\Delta V}{V}$= $(V_{MnSi}-V_{(MnFe)Si})/V_{(MnFe)Si}$.  Taking $K$=1.64 Mbar~\cite{15} and $\dfrac{\Delta V}{V}$=8.96$\cdot 10^{-3}$ (it follows from the given above the lattice parameters values), one obtain $P$=14.63 kbar. Surprisingly this value practically coincide with the pressure corresponding to the phase transition in the pure MnSi at zero temperature~\cite{1,2,3,4}. This adds an extra argument in favor of quantum criticality of (MnFe)Si in the vicinity of iron concentration 0.15\%.

\begin{figure}[htb]
\includegraphics[width=80mm]{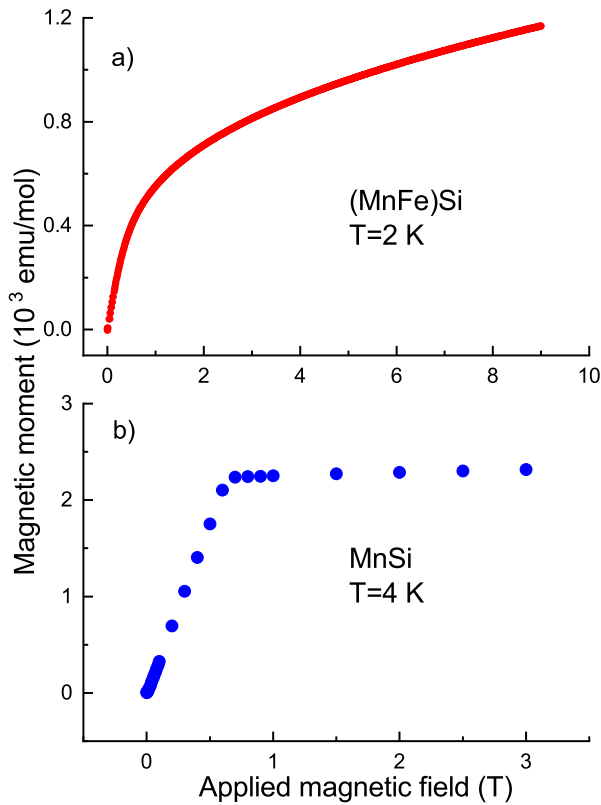}
\caption{\label{fig1} (Color online) Magnetization curves for (MnFe)Si (a) and MnSi (b)~\cite{7,17}.}
\end{figure}

\begin{figure}[htb]
\includegraphics[width=80mm]{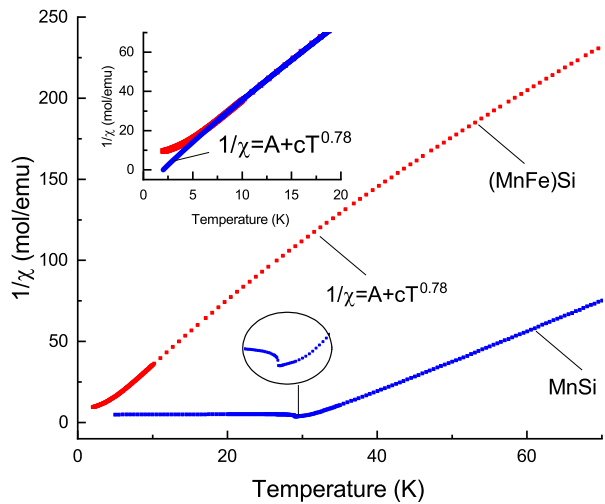}
\caption{\label{fig2} (Color online) The inverse magnetic susceptibility $1/\chi$ for (MnFe)Si and MnSi~\cite{7,17} as measured at 0.01 T.}
 \end{figure}

\section{Experimental}
We performed some magnetic, dilatometric, electrical and heat capacity measurements to characterize the sample of (MnFe)Si.  All measurements were made making use the  Quantum Design PPMS system with the heat capacity and vibrating magnetometer moduli. The linear expansion of the sample was measured by the capacity dilatometer~\cite{16}. The resistivity data were obtained with the standard four terminals scheme using the spark welded Pt wires as electrical contacts.

\begin{figure}[htb]
\includegraphics[width=80mm]{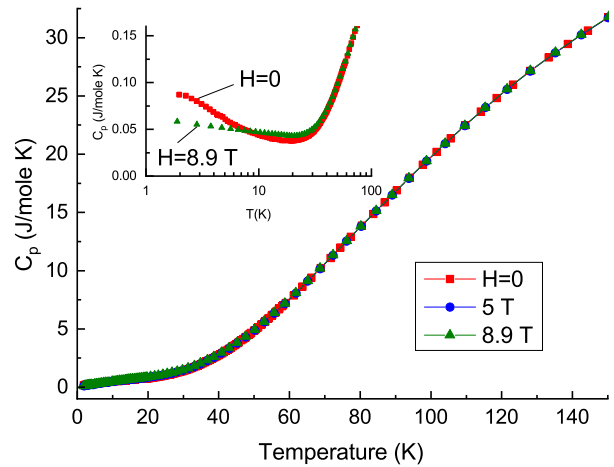}
\caption{\label{fig3} (Color online) Specific heat of (MnFe)Si as a function of temperature at different magnetic fields.  Specific heat of (MnFe)Si divided by temperature $C_p/T$ is shown in the inset in the logarithmic scale at zero magnetic field. }
\end{figure}

\begin{figure}[htb]
\includegraphics[width=80mm]{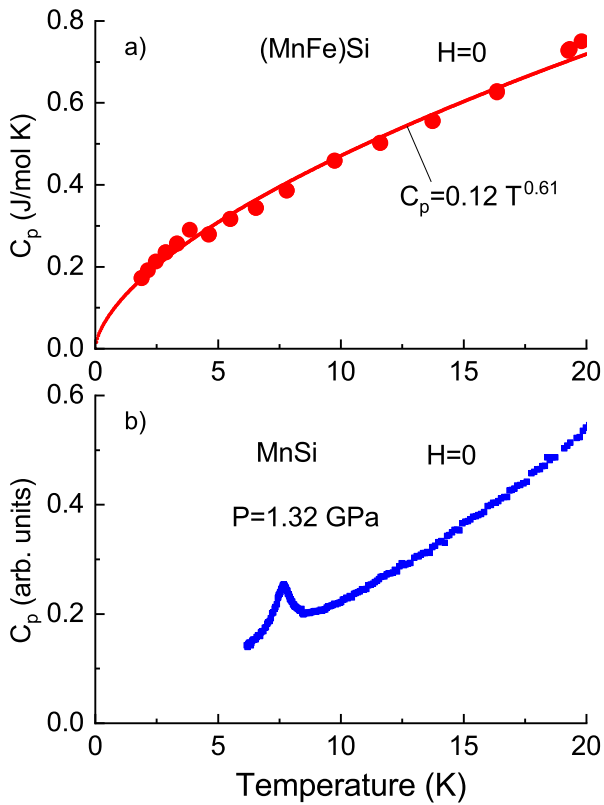}
\caption{\label{fig4} (Color online) a) Specific heat of (MnFe)Si as a function of temperature in the 2--20 K range. The line is the power function fit the experimental data (shown in the plot).
b) Specific heat of MnSi at high pressure measured by the ac-calorimetry technique~\cite{19}.}
\end{figure}

\begin{figure}[htb]
\includegraphics[width=80mm]{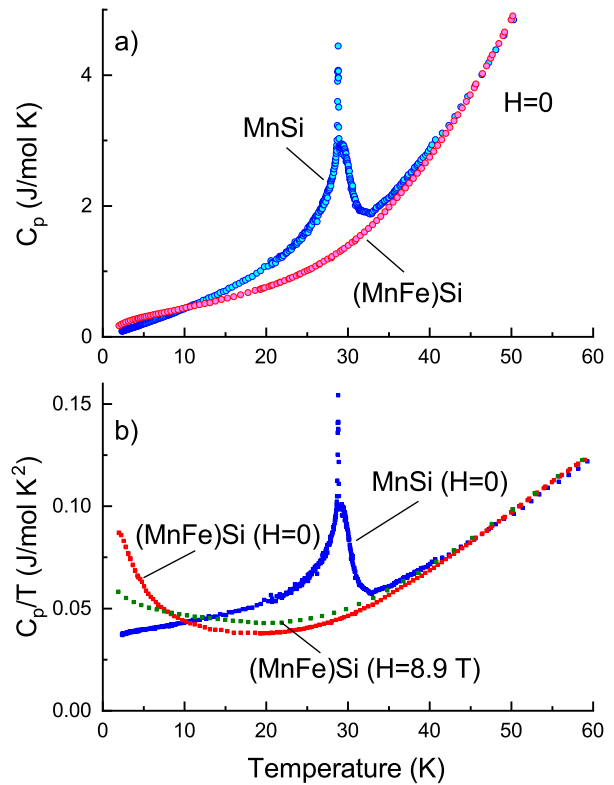}
\caption{\label{fig5} (Color online) Temperature dependence of $C_p$ (a) and $C_p/T$ (b) for (MnFe)Si and MnSi~\cite{7,20}.}
\end{figure}

\begin{figure}[htb]
\includegraphics[width=80mm]{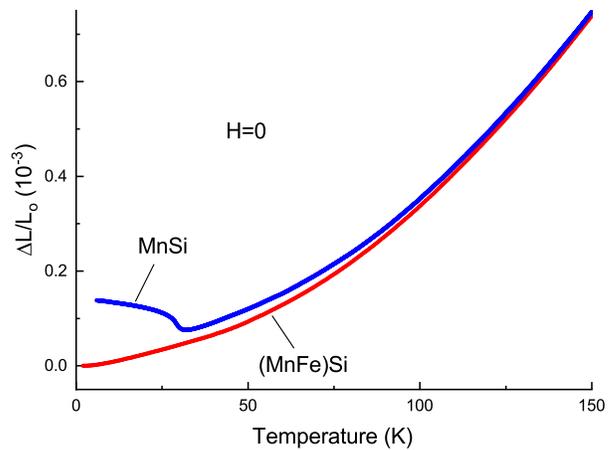}
\caption{\label{fig6} (Color online) Dependence of linear thermal expansion of (MnFe)Si and MnSi~\cite{7,20} on temperature. MnSi data reduced to (MnFe)Si ones at 200 K for better viewing.} 
\end{figure}

\begin{figure}[htb]
\includegraphics[width=80mm]{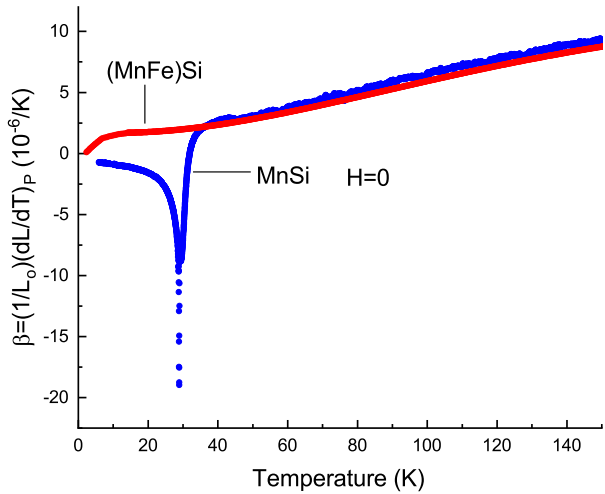}
\caption{\label{fig7} (Color online) Linear thermal expansion coefficients of (MnFe)Si and MnSi~\cite{7,20}. } 
\end{figure}

\begin{figure}[htb]
\includegraphics[width=80mm]{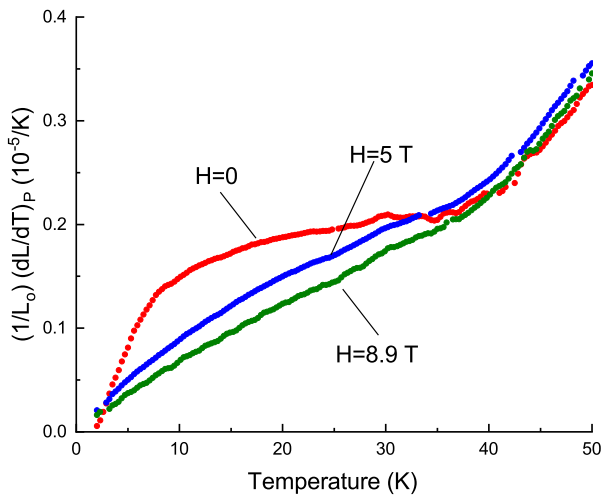}
\caption{\label{fig8} (Color online) Linear thermal expansion coefficients of (MnFe)Si as functions of temperature and magnetic fields.} 
\end{figure}

\begin{figure}[htb]
\includegraphics[width=80mm]{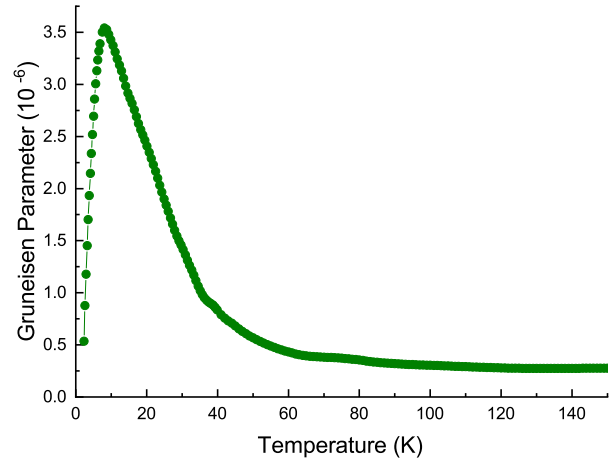}
\caption{\label{fig9} (Color online) Gruneisen ratio tends to diverse at $T\rightarrow0$. This tendency is interrupted by a peculiar behavior of the thermal expansion coefficient.} 
\end{figure}

\begin{figure}[htb]
\includegraphics[width=80mm]{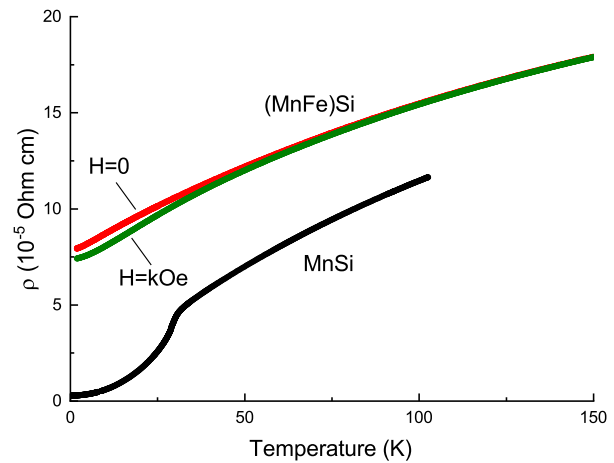}
\caption{\label{fig10} (Color online) Resistivities of (MnFe)Si and MnSi~\cite{21} as functions of temperature.} 
\end{figure}

\begin{figure}[htb]
\includegraphics[width=80mm]{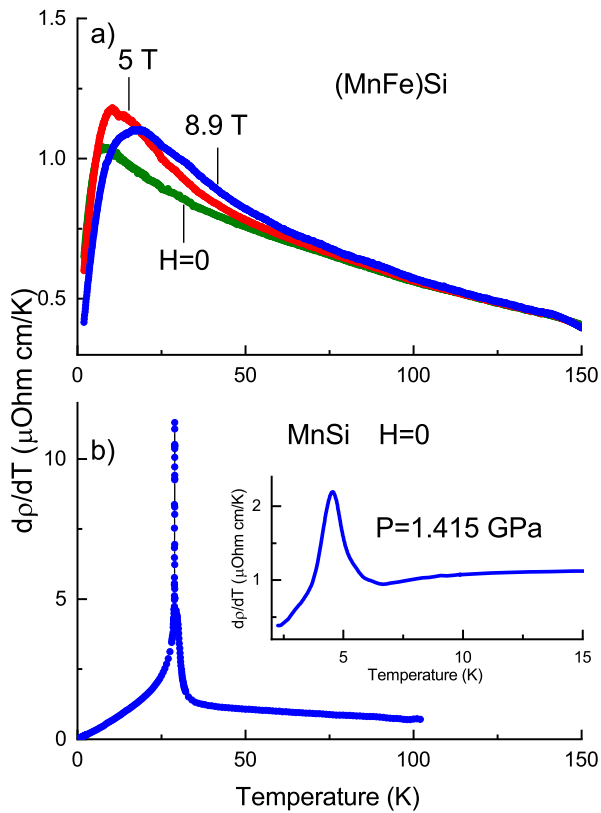}
\caption{\label{fig11} (Color online)Dependence of temperature derivative of resistivity of (MnFe)Si and MnSi on temperature: a) $d\rho/dT$ of (MnFe)Si  as functions of temperature and magnetic fields, b) $d\rho/dT$ of MnSi as a function of temperature at ambient and high pressure (in the inset)~\cite{8,21}.} 
\end{figure}

The experimental results are displayed in Fig.~\ref{fig1}--\ref{fig11}. Whenever it is possible the corresponding data for pure MnSi are depicted at the same figures to facilitate comparisons of the data.

In Fig.~\ref{fig1} the magnetization curves for both (MnFe)Si and MnSi are shown. As it follows the magnetization of (MnFe)Si (a) does not reveal an existence of the spontaneous magnetic moment in contrast with a case of MnSi.  From the saturated magnetization of MnSi at high field (Fig.~\ref{fig1}b), the magnetic moment per atom Mn is 0.4$\mu_B$.

As seen from Fig.~\ref{fig2} the magnetic susceptibility $\chi$ of (MnFe)Si does not obey the Curie-Weiss law, which clearly works  in the paramagnetic phase of MnSi. The temperature dependence of $1/\chi$ for (MnFe)Si is well described in the range 5-150 K by the expression $1/\chi=A+cT^{0.78}$, which was also observed for some substances with quantum critical behavior~\cite{18}. This expression can rewritten in the form $(1/\chi - 1/\chi_0)^{-1}=cT^{-1}$, implying a divergence of the quantity $(1/\chi - 1/\chi_0)^{-1}$. The nature of the anomalous part of the $1/\chi$ at $<5$~K  (see inset in Fig.~\ref{fig2}) will be discussed later.

As can be seen from Fig.~\ref{fig3} magnetic field does not influence much the specific heat of (MnFe)Si at least at high temperatures. Also is seen in the inset of Fig.~\ref{fig3} that the ratio of $C_p/T$ does not fit well the logarithmic law.

The power law behavior of $C_p$ in the range to 20~K is characterized by the exponent $\sim 0.6$ (Fig.~\ref{fig4}), which immediately leads to the diverging expression for $C_p/T\sim T^{-1+0.6}$ (see Fig.~\ref{fig5}b). This finding contradicts to the data~\cite{12} declaring the logarithmic divergence of $C_p/T$ for (MnFe)Si in about the same temperature range (see the inset in Fig.~\ref{fig3}). In Fig.~\ref{fig4}b is shown how the phase transition in MnSi at high pressure close to the quantum critical region influences the specific heat. The additional illustration of this kind is provided by the resistivity data (see Fig.~\ref{fig11}). So one cannot find any similar evidence in Fig.~\ref{fig4}a for the would be phase transition, which was suggested in~\cite{14}.

Fig.~\ref{fig5} shows the temperature dependences of specific heats $C_p$ (a) and specific heats divided by temperature $C_p/T$ (b) for (MnFe)Si and MnSi. As can be seen both quantities do not differ much at temperatures above the magnetic phase transitions in MnSi even with applied magnetic field.  The great difference arises at and below phase transition temperatures in MnSi. The remarkable thing is the diverging behavior of $C_p/T$ that is removed by an application of strong magnetic field (Fig.~\ref{fig5}b) leading to  the finite value of $(C_p/T)$ at T=0 corresponding to the electronic specific heat term $\gamma$, therefore restoring Fermi liquid picture

As is seen in Fig.~\ref{fig6} the magnetic phase transition in MnSi is signified by a significant volume anomaly. Nothing of this kind exists on the thermal expansion curve of (MnFe)Si. Probably a somewhat different situation can be observed in Fig.~\ref{fig7}, which displays the temperature dependences of linear thermal expansion coefficients $\beta=(1/L_0) (dL/dT)_p$ for (MnFe)Si and MnSi. It is seen a surprisingly good agreement between both data at high temperature. A specific feature of $\beta$ of (MnFe)Si is a small tail at T$<5$~K. This tale inclines to cross the temperature axis at finite value therefore tending to the negative $\beta$ as it does occur in MnSi in the phase transition region (see Figs.~\ref{fig7} and \ref{fig8}). Just this behavior of $\beta$ creates sudden drop at low temperatures in the seemingly diverging ratio $\beta/C_{p}$, which conditionally may be called the Gruneisen parameter (See Fig.~\ref{fig9}). 
 

Fig.~\ref{fig8} shows that magnetic field strongly influences the "tale" region of the thermal expansion coefficient of (MnFe)Si that indicates its fluctuation nature. This feature should be linked to the anomalous part of the $1/\chi$ at $<5$~K  (Fig.~\ref{fig2}).

Resistivities of (MnFe)Si and MnSi as functions of temperature are shown in Fig.~\ref{fig10}. The quasi linear non Fermi liquid behavior of resistivity of (MnFe)Si at low temperature in contrast with the MnSi case is quite obvious. With temperature increasing the resistivity of (MnFe)Si evolves to the "saturation" curve typical of the strongly disordered metals  and similar to the post phase transition branch of the resistivity curve of MnSi~\cite{22}.

A comparison of Fig.~\ref{fig11} (a) and (b) shows a drastic difference in behavior of $d\rho/dT$ at the phase transition in MnSi and in (Mn,Fe)Si in the supposedly critical region. The peculiar form of $d\rho/dT$ of (Mn,Fe)Si does not look as a phase transition feature though it certainly reflects an existence of significant spin fluctuations. This feature should be related to the anomalies of the magnetic susceptibility (fig.~\ref{fig2}) and thermal expansion coefficient (fig.~\ref{fig8}).

\section{Discussion}
As we have shown in the Introduction the lattice parameter of our sample of (MnFe)Si corresponds to the one of compressed pure MnSi by pressure about 1.5 GPa. At this pressure and zero temperature the quantum phase transition in MnSi does occur, nature and properties of which are still under discussion~\cite{7}. Alternative way to reach the quantum regime is to use so called "chemical pressure" doping MnSi with suitable "dopants" that could avoid disturbing inhomogeneous stresses  arising at conventional pressure loading. So as it appeared the composition Mn$_{0.85}$Fe$_{0.15}$Si indeed demonstrated properties typical of the quantum critical state ~\cite{11,12}. However the conclusions of Ref.~\cite{11,12} were disputed in the publication~\cite{14}, authors of which claim on the basis of the muon spin relaxation experiments that 15\% Fe-substituted (Mn,Fe)Si experiences a second order phase transition at ambient pressure then reaching a quantum critical point at pressure $\sim$21--23 kbar.

With all that in mind we have carried out a number of measurements trying to elucidate the problem. Below we summarize our finding.

\begin{enumerate}
\item There is no spontaneous magnetic moment in (MnFe)Si at least at 2~K (Fig.1). Magnetic susceptibility of (MnFe)Si can be described by the expression $1/\chi=A+cT^{0.78}$  or $(1/\chi - 1/\chi_0)^{-1} = cT^{-1}$ in the temperature range $\sim$5--150~K, implying divergence of the quantity $(1/\chi - 1/\chi_0)^{-1}$. This behavior also was observed earlier in case of some substances close to quantum critical region (Fig.~\ref{fig2})~\cite{18}. At $T<5$~K a behavior of $1/\chi$ deviates from the mentioned expression in a way, which can be traced to the analogous feature at the fluctuation region of MnSi at $T>T_c$ (see the round inset in Fig.~\ref{fig2}).
\item Specific heat of (MnFe)Si is well defined by the simple power expression $C\sim T^{0.6}$ in the range 2--20~K , which does not show any features inherited to phase transitions as it take place in case of MnSi at pressure close to the quantum phase transition (Fig.~\ref{fig3},\ref{fig4}). This expression immediately leads to the divergence of the quantity $C_p/T\sim T^{-1+0.6}$, which can be suppressed by magnetic field that leads to restoring Fermi liquid picture with finite value of electronic specific heat term $\gamma$ (Fig.~\ref{fig5}).
\item The thermal expansion experiment with (MnFe)Si does not reveal any features that can be linked to a phase transition (Fig.~\ref{fig6}). However the thermal expansion coefficient $\beta$ show a low temperature tale, which inclines to cross the temperature axis at finite value tending to become negative as it occurs in MnSi (Fig.~\ref{fig7},\ref{fig8}). This specifics of $\beta$ causes a sudden low temperature drop of the Gruneisen parameter otherwise it would diverge at $T\rightarrow 0$.  An application of magnetic field suppresses this kind of behavior of the thermal expansion coefficient therefore revealing its fluctuation nature (Fig.~\ref{fig8}).
\item The resistivity of (MnFe)Si clearly demonstrates non Fermi liquid behavior with no specifics indicating a phase transition. However, the temperature derivative of resistivity $d\rho/dT$ of (MnFe)Si shows non trivial form, which indicates an existence of significant spin fluctuations. That should be related to the low temperature "tales" both magnetic susceptibility and thermal expansion coefficient.
\end{enumerate}
\section{Conclusion}
Finally, magnetic susceptibility  in the form $(1/\chi - 1/\chi_0)^{-1}$ and Gruneisen parameter $\beta/C_p$ in (MnFe)Si show diverging behavior, which is interrupted at about 5~K by  factors linked somehow with spin fluctuations analogues to ones preceding the phase transition in MnSi (see Fig.~\ref{fig2},\ref{fig7},\ref{fig8}). Specific heat divided by temperature $C_p/T$ of (MnFe)Si clearly demonstrate diverging behavior to 2~K.  The electrical resistivity of (MnFe)Si exhibits non Fermi liquid character.

General conclusions: there are  no thermodynamic evidences in favor of a second order phase transition for the 15\% Fe substituted MnSi. The trajectory corresponding to the present composition of (MnFe)Si is a critical one, i.e. approaching quantum critical point at lowering temperature, which agrees with conclusions made in Ref.~\cite{11,12}. However, the critical trajectory in fact is a tangent to the phase transition line and therefore some properties inevitably would be influenced by the cloud of spin helical fluctuations bordering the phase transition. This situation produces some sort of a mixed state instead of a pure quantum critical one that probably was seen in the experiments~\cite{14}.

\section{Acknowledgements}
We express our gratitude to I.P. Zibrov and N.F. Borovikov for technical assistance.
AEP and SMS greatly appreciate financial support of the Russian Foundation for Basic Research (grant No. 18-02-00183), the Russian Science Foundation (grant 17-12-01050).

\end{document}